\documentclass[conference]{IEEEtran}
\IEEEoverridecommandlockouts
\usepackage{cite}

\usepackage{amsmath,amssymb,amsfonts}
\usepackage{algorithmic}
\usepackage{algorithm}
\usepackage{graphicx}
\usepackage{subfig}
\usepackage{textcomp}
\usepackage{xcolor}
\usepackage{multirow}

\usepackage{epstopdf}
\def\BibTeX{{\rm B\kern-.05em{\sc i\kern-.025em b}\kern-.08em
    T\kern-.1667em\lower.7ex\hbox{E}\kern-.125emX}}
\begin{document}

\title{Dynamic Quality-Latency Aware Routing for LLM Inference in Wireless Edge-Device Networks\\
}
\author{
    Rui Bao\IEEEauthorrefmark{1},
    Nan Xue\IEEEauthorrefmark{1},
    Yaping Sun\IEEEauthorrefmark{2}\IEEEauthorrefmark{3},
    Zhiyong Chen\IEEEauthorrefmark{1} \\
    \IEEEauthorrefmark{1}Cooperative Medianet Innovation Center, Shanghai Jiao Tong University, Shanghai 200240, China \\
    \IEEEauthorrefmark{2}Department of Broadband Communication, Pengcheng Laboratory, Shenzhen 518000, China \\
    \IEEEauthorrefmark{3}Future Network of Intelligence Institute (FNii), CUHK (Shenzhen), Shenzhen 518172, China \\
    Emails: \{851756936, nan.xue, zhiyongchen\}@sjtu.edu.cn, sunyp@pcl.ac.cn

\thanks{
    This work is supported in part by NSF of China under Grant 62222111 and Grant 62431015, and in part by the Science and Technology Commission Foundation of Shanghai under Grant 24DP1500702.
}
}

\maketitle

\begin{abstract}
The integration of wireless communications and Large Language Models (LLMs) is poised to unlock ubiquitous intelligent services, yet deploying them in wireless edge-device collaborative environments presents a critical trade-off between inference quality and end-to-end latency. A fundamental mismatch exists between task complexity and resource allocation: offloading simple queries invites prohibitive latency, while on-device models lack the capacity for demanding computations. To address this challenge, we propose a dynamic, quality-latency aware routing framework that orchestrates inference between a lightweight model on the mobile device and a powerful model on the edge server. Our framework employs two distinct cost models: for single-turn queries, it fuses a BERT-predicted semantic score with communication and computation overheads; for multi-turn dialogues, it further quantifies context-aware costs arising from model switching and KV-cache management. While maintaining full inference quality, extensive experiments demonstrate that our framework cuts average response latency by 5-15\% and reduces large model invocations by 10-20\% against competitive baselines on MMLU, GSM8K, and MT-Bench-101 benchmarks.
\end{abstract}


\section{Introduction}
Large Language Models (LLMs) are rapidly transforming human interaction with the digital world, serving as powerful intelligent assistants across a multitude of applications. Central to this vision is the rise of on-device LLMs, which promise low-latency responses, enhanced privacy, and offline accessibility. Small language models (SLMs), such as Llama-3-8B, Phi-3-mini, and Qwen2.5-3B, demonstrate remarkable performance on a wide range of common tasks \cite{grattafiori2024llama, abdin2024phi, qwen2.5}. However, their capabilities in complex, multi-step reasoning still lag significantly behind their larger counterparts. The fundamental barrier lies in the immense computational and memory resources demanded by high-fidelity reasoning models, resources that far exceed the capacity of today's resource-constrained mobile devices. This inherent resource-capability conflict necessitates a shift towards an edge-device collaborative architecture, leveraging the best of both worlds to deliver truly intelligent on-device experiences.

For edge-device collaborative LLM inference, one prevailing paradigm is model partitioning. For instance, the Wireless Distributed Mixture-of-Experts (WDMoE) architecture decomposes MoE layers by placing the gating network on an edge server and distributing the individual expert networks among mobile devices \cite{xue2024wdmoe}. Similarly, \cite{zhang2024edgeshard} devises a more general framework to partition an LLM into shards and deploy them on distributed devices, dynamically optimizing device selection and partition points. However, this partitioning paradigm faces two significant challenges in the wireless context. First, these methods are architected around a single, large-scale model, leading to unnecessary computational overhead for simple queries. Second, they require transmitting large intermediate tensors between the device and edge, which consumes substantial wireless bandwidth and introduces significant latency. While physical layer innovations like Channel Denoising Diffusion Models (CDDM) can enhance transmission robustness by mitigating channel noise \cite{wu2024cddm}, they do not resolve the fundamental issue of high latency caused by frequent, high-volume data exchange.

Another prominent paradigm is speculative sampling, which accelerates the inference of a large target model at the token-level by generating multiple tokens from each model call \cite{chen2023accelerating}. In this approach, a smaller, faster draft model first generates a sequence of candidate tokens, which are then validated in parallel by the larger target model \cite{chen2023accelerating, leviathan2023fast}. A key advantage of this method is that a novel rejection sampling scheme ensures the final output is mathematically identical to what the larger model would have produced alone, preserving its exact distribution without compromising quality.  This concept of token-level collaboration has also been explicitly adapted for the edge-cloud context, where an on-device SLM interacts at the token-level with a cloud-based LLM to achieve LLM-comparable quality at a significantly lower cost \cite{hao2024hybrid}. However, this paradigm introduces significant new overheads: the frequent, token-level communication required for validation can be costly over a wireless link, while concurrently running and loading both a draft and a target model imposes a substantial memory and computational burden on resource-constrained devices.

In the domain of large and small models collaboration, \cite{Ong2024RouteLLM} proposes a paradigm for collaboration between a strong and a weak model through training a router on human preference data to dynamically select models. \cite{Ding2024Hybrid} introduces a router that makes decisions based on predicted query difficulty and a user-defined quality level that can be tuned at test time. More advanced systems like MixLLM \cite{wang2025mixllm} employ a contextual-bandit-based approach to manage the complex tradeoffs between quality, cost, and latency for a varying set of candidate LLMs. This collaborative paradigm of routing between large and small models is inherently well-suited for the edge-device collaboration. However, directly applying the paradigm to edge-device collaboration reveals two gaps. First, their cost models often fail to account for the stochastic latency inherent in wireless networks. Second, and more critically, they overlook the substantial overhead introduced by Key-Value (KV) cache management (i.e., recomputation or transmission) within multi-turn dialogues \cite{gao2024cost, pope2023efficiently}.

To bridge these gaps, we propose a \textit{Dynamic Quality-Latency Aware Routing Framework} specifically designed for wireless edge-device LLM collaboration. Further, we devise a novel fusion mechanism that integrates semantic difficulty prediction with a fine-grained, end-to-end latency model. For semantic prediction, we employ a lightweight BERT router inspired by RouteLLM \cite{Ong2024RouteLLM}, while our custom latency model explicitly accounts for both on-device computation and wireless communication costs. We introduce a context-length-aware routing strategy for multi-turn dialogues. To the best of our knowledge, this is the first routing framework designed for wireless edge-device collaboration that quantitatively models the KV-cache recomputation overhead during model switching, enabling efficient and fluid conversations. Experiment results showcase the effectiveness of our framework.

\section{System Model and Problem Formulation}
\label{sec:system_model}

\subsection{Collaborative Inference Architecture}
Our framework operates within a wireless edge-device collaborative architecture, designed to leverage the complementary strengths of local and remote computation. The system comprises three main components:
\begin{itemize}
    \item \textbf{Mobile Device:} A resource-constrained device, such as a smartphone, that hosts a small language model (SLM), e.g., Qwen2.5-3B \cite{qwen2.5}. The SLM is optimized for low-latency inference on simple queries.
    \item \textbf{Edge Server:} A powerful server equipped with high-performance GPUs, hosting a large language model (LLM), e.g., QwQ-32B \cite{qwen2.5}. The LLM provides high-quality responses for complex reasoning tasks.
    \item \textbf{Wireless Link:} The mobile device and edge server are connected via a wireless channel characterized by a bandwidth of $B$ and a fixed transmission overhead of $\delta$. To align with the data format of LLMs, we measure the bandwidth $B$ in tokens per second (tokens/s). This rate is derived from the physical layer's bit rate (e.g., Mbps or Gbps in 5G/Wi-Fi 6) by assuming an average number of bits per token, which accounts for character encoding and average word length.
\end{itemize}

\begin{figure}[t]
\centering
\includegraphics[width=\columnwidth]{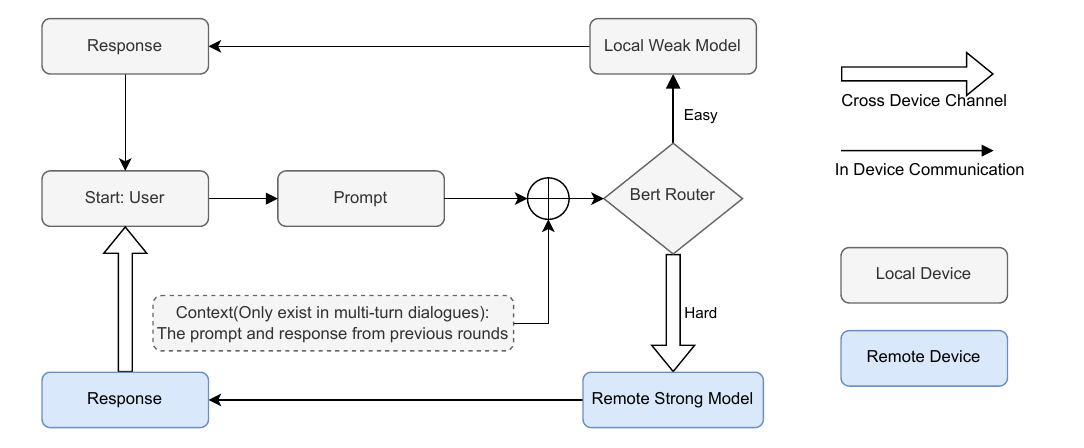}
\caption{The unified workflow of the dynamic routing framework. The context (dotted box) is considered only in multi-turn dialogues.}
\label{fig:workflow}
\end{figure}

The overall workflow is illustrated in Fig. \ref{fig:workflow}. Upon receiving a user prompt, a lightweight router on the end device decides whether to process the request locally using the SLM or to offload it to the edge server for processing by the LLM. For multi-turn dialogues, the router's decision is also informed by the accumulated conversation history, or context, which is represented by the dotted box in the figure.

\begin{figure*}[t]
\centering
\includegraphics[width=0.87\textwidth]{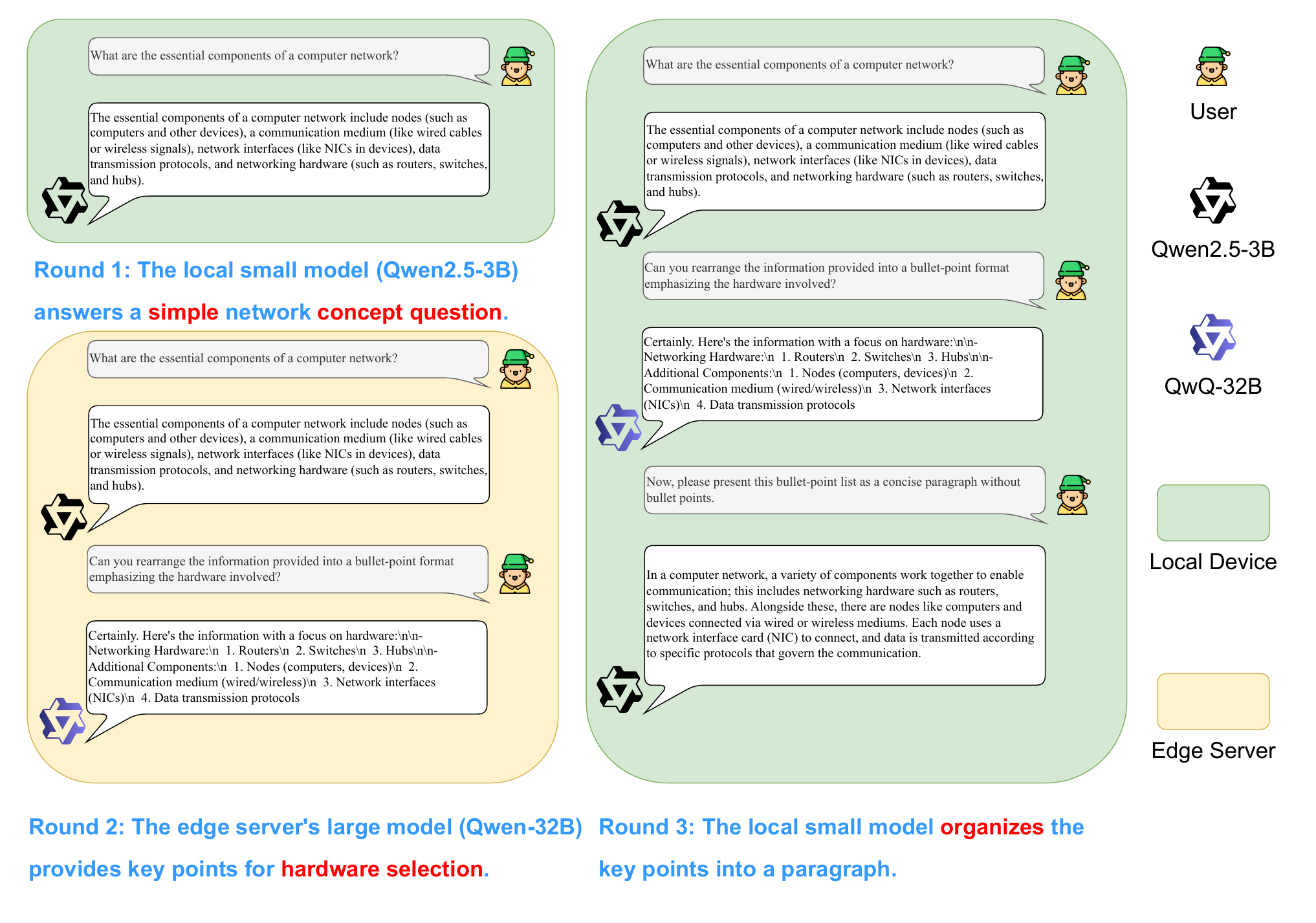}
\caption{An example of a three-turn dialogue showcasing the dynamic routing process. The router intelligently switches between the local SLM and the edge LLM based on the query's complexity and the accumulated conversation context, optimizing for both latency and response quality.}
\label{fig:dialogue_example}
\end{figure*}

\subsection{Latency Cost Model}
To make informed routing decisions, we establish a fine-grained latency model. Let $L_p$ be the token length of the current prompt and $C_t$ be the token length of the accumulated context from previous turns. For single-turn queries, $C_t = 0$. 
We define the per-token computation times for the SLM and the edge LLM as $\gamma_s$ and $\gamma_e$ respectively. The router's per-token processing time is denoted by $\tau_r$. Finally, we denote the average response lengths from the SLM and the edge LLM as $\bar{L}_r^s$ and $\bar{L}_r^e$ respectively.

\paragraph{Computation Latency $T_{\text{comp}}$} is the time taken to process the input tokens (prefill stage). If the same model is used as in the previous turn, only the new prompt needs to be processed. If the model is switched, the context from the previous turn must be re-processed.
\begin{equation}
T_{\text{comp}}^{k} = \gamma_k ( \Delta C_t + L_p ), \quad k \in \{s, e\}
\end{equation}
where $\Delta C_t$ is the length of the context to be re-prefilled. $\Delta C_t > 0$ only when switching models in multi-turn dialogues; otherwise, $\Delta C_t = 0$.

\paragraph{Communication Latency $T_{\text{comm}}$} is incurred only when offloading to the edge server. It includes the time to upload the prompt and context, and to download the response. The term $\delta$ represents a fixed transmission overhead, accounting for message-size-independent delays such as signal propagation and protocol handshaking. The total communication latency is thus:

\begin{align}
T_{\text{comm}}^{s} &= 0 \\
T_{\text{comm}}^{e} &= \underbrace{\frac{C_t + L_p}{B}}_{\text{Upload}} + \delta + \underbrace{\frac{\bar{L}_r^e}{B}}_{\text{Download}}
\end{align}

\paragraph{Router Latency $T_{\text{router}}$} is the time the on-device router takes to make a decision.
\begin{equation}
T_{\text{router}} = \tau_r (C_t + L_p)
\end{equation}

\paragraph{Generation Latency $T_{\text{gen}}$} is the time taken to generate the response tokens (decode stage).
\begin{equation}
T_{\text{gen}}^{k} = \gamma_k \bar{L}_r^k, \quad k \in \{s, e\}
\end{equation}

The latency for choosing the SLM $T_{\text{total}}^{s}$, the LLM $T_{\text{total}}^{e}$ and the total end-to-end latency for a query, $T_{\text{total}}$ can be expressed as:
\begin{align}
T_{\text{total}}^{s} &= T_{\text{comp}}^{s} + T_{\text{router}} + T_{\text{gen}}^{s} \\
T_{\text{total}}^{e} &= T_{\text{comp}}^{e} + T_{\text{comm}}^{e} + T_{\text{router}} + T_{\text{gen}}^{e} \\
T_{\text{total}} &= d_t \cdot T_{\text{total}}^{s} + (1-d_t) \cdot T_{\text{total}}^{e}
\end{align}
where $d_t \in \{0, 1\}$ represents selection between edge server and mobile device.

\subsection{Problem Formulation}
We formulate an optimization problem with the aim of minimizing the total latency while maintaining service quality (e.g., response accuracy or score). Routing policy $\pi$ determines the model selection $d_t$ for each incoming query $q_t$ at turn $t$. The problem is formulated as:
\begin{equation}
\min_{\pi} \quad \mathbb{E}[T_{\text{total}}(d_t)]
\end{equation}
\begin{equation*}
\label{eq:constraint}
\text{s.t.} \quad \mathbb{E}[\mathcal{Q}(d_t)] \ge Q_{\text{target}}
\end{equation*}
where $T_{\text{total}}(d_t)$ is the latency corresponding to decision $d_t$, $\mathcal{Q}(d_t)$ represents the quality score (e.g., accuracy) for that decision, and $Q_{\text{target}}$ is the minimum acceptable quality threshold. This formulation captures the fundamental latency-quality tradeoff inherent in the system.

\section{Dynamic Quality-Latency Aware Routing}

\label{sec:routing_framework}
To address the optimization problem formulated in Section \ref{sec:system_model}, we propose a dynamic routing framework that intelligently arbitrates between the local SLM and the edge LLM. The framework's decision logic consists of two main stages: a semantic-aware router that predicts the query's complexity and a fusion mechanism that combines this prediction with our latency cost model to make a final, quality-latency aware decision.

\subsection{Semantic-Aware Router}
To assess the semantic difficulty of an incoming request, our framework directly employs the pre-trained, lightweight BERT router in RouteLLM \cite{Ong2024RouteLLM}. This component functions as a foundational module in our system by generating an initial, quality-driven probability score, which indicates whether a query requires the powerful edge LLM.

Functionally, for a given turn $t$, the router takes the current prompt $q_t$ and the accumulated dialogue history $\mathrm{Context}_t$ as input. The combined input sequence is fed into the BERT model, and the resulting context vector, $\mathbf{h}_{\mathrm{CLS}}$, is concatenated with the prompt length $L_p$ and passed through a linear layer with a sigmoid activation function to produce the probability score $p_t \in [0, 1]$:
\begin{equation}
p_t = \sigma\bigl(\mathbf{W}^{\mathsf T}[\mathbf{h}_{\mathrm{CLS}}; L_p] + b\bigr)
\label{eq:bert_prob}
\end{equation}
where $\mathbf{W}$ and $b$ are the parameters of the router's pre-trained classification head. This score, $p_t$, serves as the primary input to our novel quality-latency fusion mechanism, which is the core of our framework's decision logic.

\subsection{Quality-Latency Fusion for Routing Decisions}
While the initial probability $p_t$ captures the query's semantic complexity, it does not account for the system's operational cost. To achieve a true quality-latency tradeoff, we introduce a novel fusion mechanism that adjusts $p_t$ based on the latency difference between the two execution paths.

We compute an adjusted probability score, $\tilde{p}_t$, by penalizing the initial score with the modeled latency gap from Section \ref{sec:system_model}:
\begin{equation}
\tilde{p}_t = p_t - \alpha \cdot \bigl(T_{\text{total}}^{e} - T_{\text{total}}^{s}\bigr)
\label{eq:fusion}
\end{equation}
Here, $\alpha$ is a tunable hyperparameter that controls the weight of the latency penalty. A higher $\alpha$ makes the router more sensitive to latency, prioritizing faster responses, while a lower $\alpha$ prioritizes inference quality as predicted by $p_t$.

The final routing decision, $d_t$, is then made by comparing this fused score against a configurable threshold, $\theta \in [0, 1]$:
\begin{equation}
d_t =
\begin{cases}
    0, & \text{if } \tilde{p}_t > \theta \\
    1, & \text{if } \tilde{p}_t \le \theta
\end{cases}
\label{eq:decision}
\end{equation}
By tuning the threshold $\theta$, system operators can navigate the tradeoff curve between accuracy and latency, selecting an operating point that best suits the specific requirements of the 6G application, from highly responsive chatbots to high-fidelity reasoning engines. This mechanism provides a flexible and powerful tool for managing LLM inference in dynamic wireless environments.

\subsection{Illustrative Example}
To provide a qualitative understanding of our framework in action, Fig. \ref{fig:dialogue_example} illustrates a three-turn dialogue where the router makes dynamic decisions. The system initially uses the local SLM for a simple concept query (Turn 1), switches to the edge LLM for a complex selection request (Turn 2), and then efficiently returns to the local SLM for a format organizing task, leveraging incremental prefill for the new context (Turn 3). This showcases the framework's ability to adapt its strategy based on both query difficulty and conversation history.

\section{EXPERIMENT RESULTS}
\label{sec:experiments}

\begin{table}[t]
\centering
\caption{Key Parameters for the Latency Cost Model.}
\label{tab:params}
\renewcommand{\arraystretch}{1.2}
\begin{tabular}{l|c}
\hline
\textbf{Parameter} & \textbf{Value} \\
\hline
SLM Computation Time ($\gamma_s$) & 0.04 s/token \\
LLM Computation Time ($\gamma_e$) & 0.02 s/token \\
Wireless Bandwidth ($B$) & $2 \times 10^7$ tokens/s \\
Communication Overhead ($\delta$) & 0.02 s \\
Latency Weight ($\alpha$) & 0.03 (MMLU), 0.05 (GSM8K) \\
\hline
\end{tabular}
\end{table}

\subsection{Experimental Setup}
We conduct a series of simulated experiments to validate our proposed framework. The key parameters for our latency cost model, detailed in Table \ref{tab:params}, were configured to reflect a realistic wireless edge-device environment. Specifically, the on-device computation time for the SLM $\gamma_s$ is derived from recent large language model performance benchmarks on mobile platforms \cite{xiao2024large}. The computation time for the edge LLM $\gamma_e$ and other network parameters are informed by NVIDIA's official inference benchmarking documentation \cite{NVIDIA2024NIM} and typical 5G network characteristics.

\noindent\textbf{Models and Benchmarks:} Our setup features a local Small Language Model (SLM), Qwen2.5-3B, and a powerful edge Large Language Model (LLM), QwQ-32B \cite{qwen2.5}. We evaluated the framework on two distinct scenarios: 
\begin{itemize}
    \item \textbf{Single-Turn Queries:} We used the MMLU \cite{hendryckstest2021} benchmark for broad, multi-domain knowledge and reasoning, and GSM8K \cite{Cobbe2021GSM8K} for mathematical problem-solving.
    \item \textbf{Multi-Turn Dialogues:} We used MT-Bench-101 \cite{Bai2024MTB101}, a comprehensive benchmark for evaluating conversational capabilities across multiple turns.
\end{itemize}

\noindent To simplify our latency model and to avoid the overhead of a specific tokenizer, all experiments were conducted on English datasets, where token counts ($L_p, C_t$) are approximated by the number of words (i.e., splitting by spaces).

\noindent\textbf{Baselines:} We compare our proposed \textit{BERT Router} against a \textit{Random Router} baseline to isolate the benefits of semantic-aware routing. The Random Router replaces the semantic score $p_t$ from the BERT model (Eq. \ref{eq:bert_prob}) with a probability drawn uniformly from $\mathcal{U}(0,1)$. Crucially, this random probability is still processed through our quality-latency fusion logic (Eq. \ref{eq:fusion}), meaning the routing decision is still penalized by the modeled latency costs. This ensures a fair comparison where the primary difference is the semantic awareness of the initial score. As this baseline involves no complex computation, its own processing time $T_{\text{router}}$ is considered to be zero.

\begin{table}[t]
\centering
\caption{Baseline Model Performance on Single-Turn Benchmarks.}
\label{tab:single_turn_perf}
\renewcommand{\arraystretch}{1.2}
\begin{tabular}{llcc}
\hline
\textbf{Benchmark} & \textbf{Model} & \textbf{Acc. (\%)} & \textbf{Resp. Len.(token)} \\
\hline
\multirow{2}{*}{MMLU} & SLM (Qwen2.5-3B) & 65.79 & 1254.4 \\
& LLM (QwQ-32B) & 85.24 & 4567.9 \\
\hline
\multirow{2}{*}{GSM8K} & SLM (Qwen2.5-3B) & 73.09 & 136.3 \\
& LLM (QwQ-32B) & 89.01 & 460.1 \\
\hline
\end{tabular}
\end{table}

\subsection{Single-Turn Query Routing Performance}

\begin{figure}[t]
\centering
\includegraphics[width=\columnwidth]{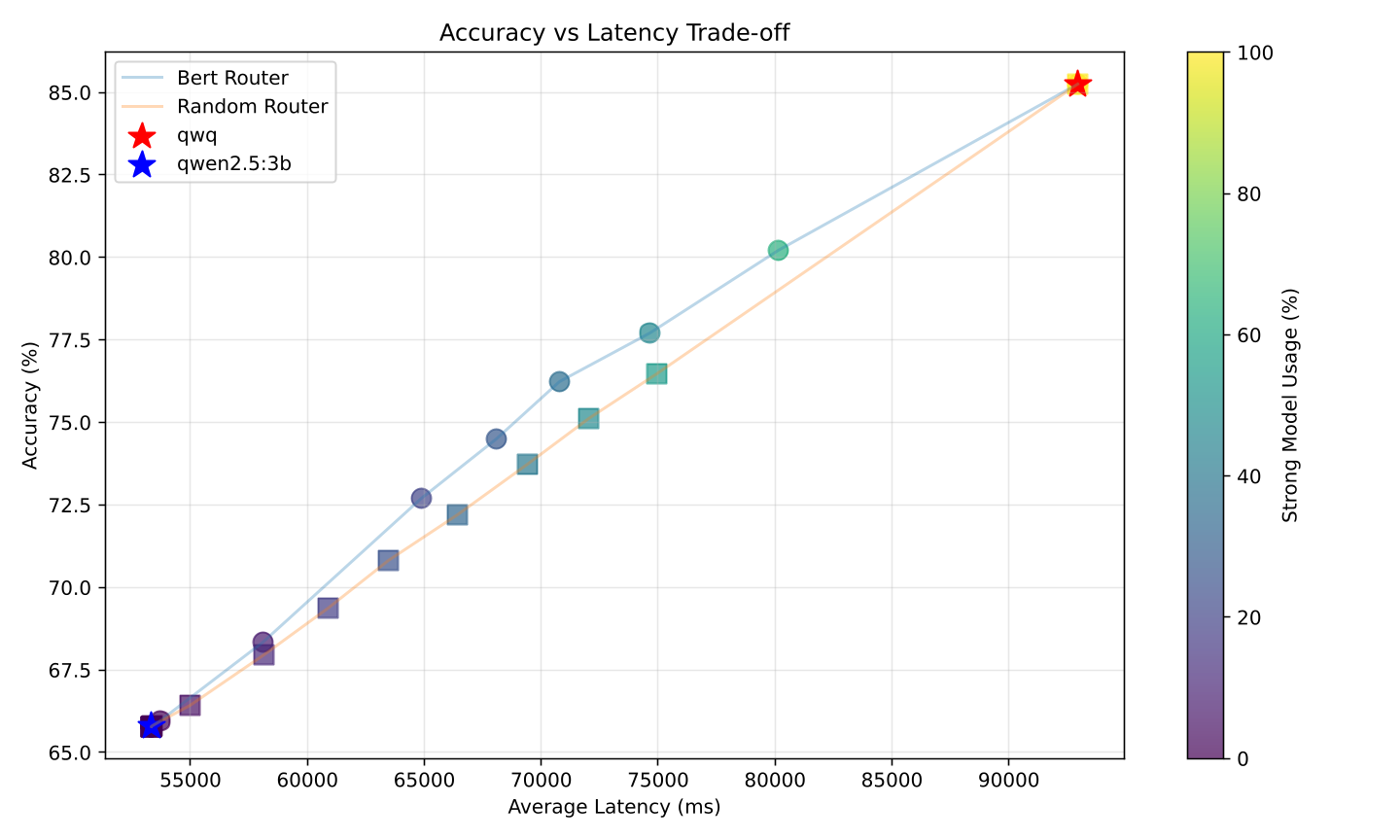}
\caption{Accuracy vs. Latency tradeoff on the MMLU benchmark. Our BERT Router achieves a better performance-latency profile.}
\label{fig:mmlu_tradeoff}
\end{figure}

On the broad-domain MMLU benchmark, our framework demonstrated significant advantages. As shown in Fig. \ref{fig:mmlu_tradeoff}, the curve for our BERT Router is consistently superior to the Random Router, indicating that for any given latency budget, our framework achieves higher accuracy. For instance, to achieve a target accuracy of 76\%, our proposed router requires an average latency of 70,784ms while invoking the expensive edge LLM on only 35\% of the queries. In contrast, the Random Router needs a higher latency of 75,039ms and a much larger 54\% LLM usage rate to attain the same accuracy level. This demonstrates the effectiveness of fusing semantic difficulty prediction with our latency cost model to achieve superior resource efficiency and responsiveness.

\begin{figure}[t]
\centering
\includegraphics[width=\columnwidth]{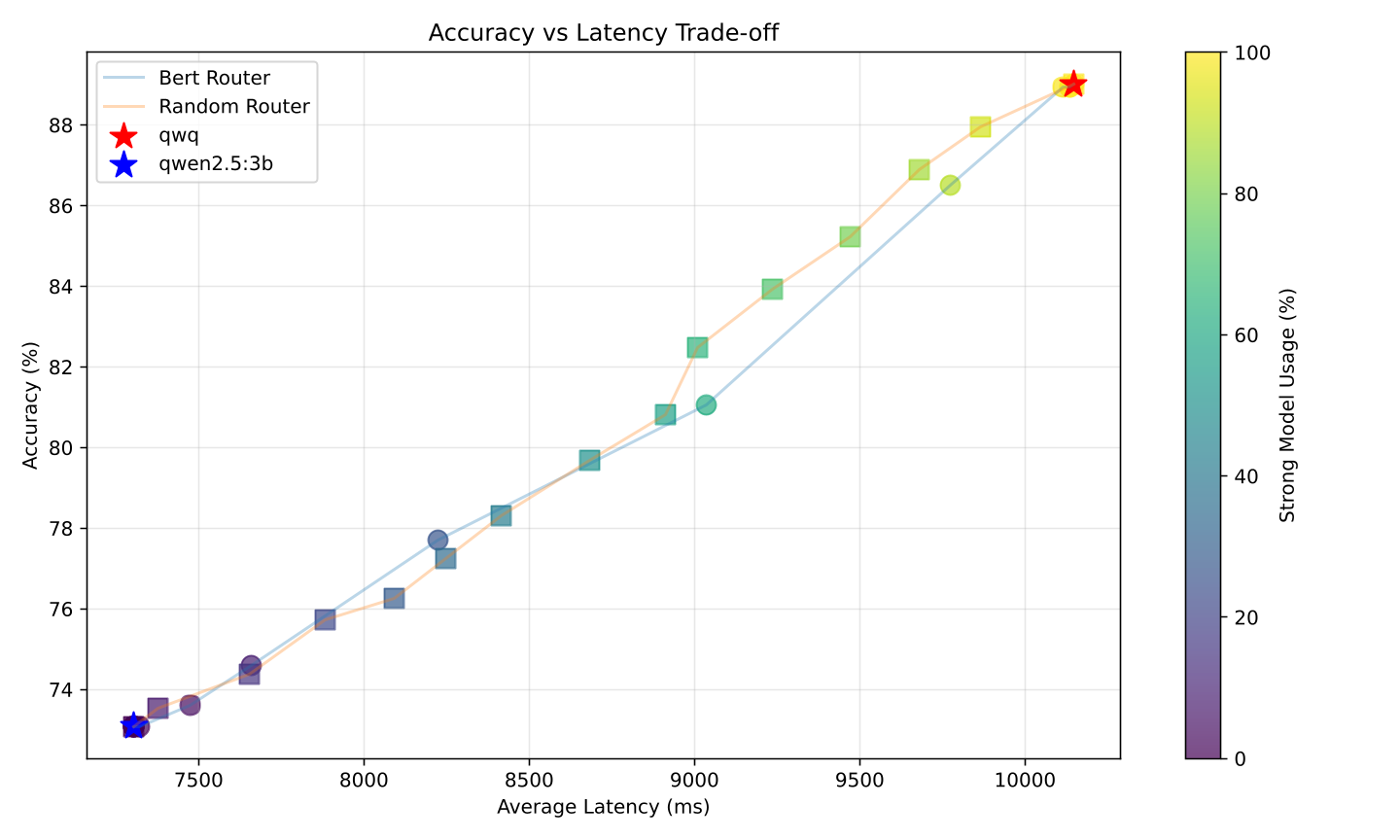}
\caption{Accuracy vs. Latency tradeoff on the specialized GSM8K benchmark. The advantage of the BERT Router is limited to the lower-accuracy region.}
\label{fig:gsm8k_latency}
\end{figure}

\begin{figure}[t]
\centering
\includegraphics[width=\columnwidth]{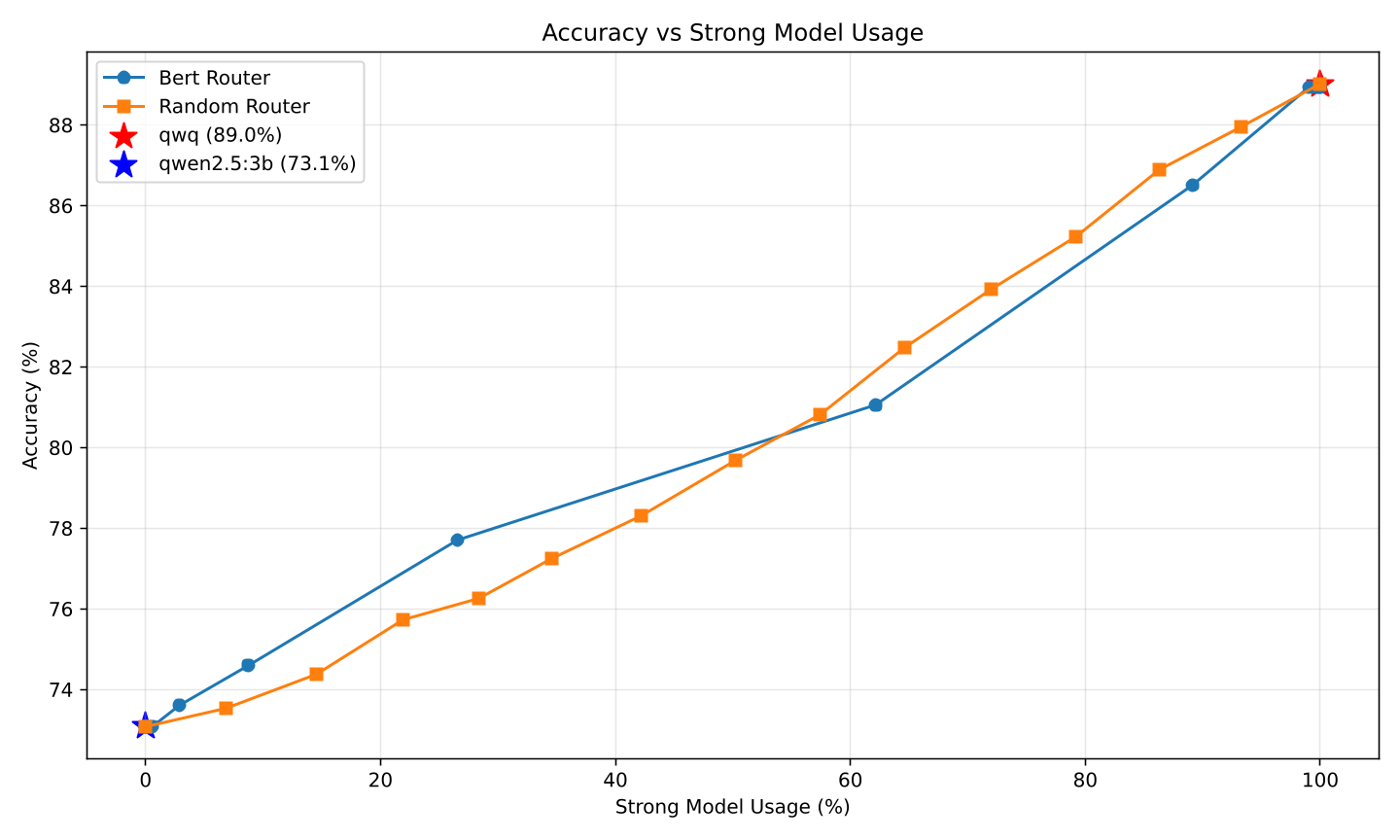}
\caption{Accuracy vs. LLM Usage on the GSM8K benchmark. This curve isolates the router's classification performance, showing it is not significantly more efficient than a random baseline at high accuracy levels.}
\label{fig:gsm8k_usage}
\end{figure}

However, the results on the specialized GSM8K benchmark reveal an important nuance. As shown in the accuracy-latency tradeoff in Fig. \ref{fig:gsm8k_latency}, while our BERT Router outperforms the baseline in the lower-accuracy region, its advantage diminishes and even reverses as the target accuracy increases. The reason for this is elucidated by the accuracy vs. LLM usage curve in Fig. \ref{fig:gsm8k_usage}. This curve, which isolates the router's classification capability from latency effects, shows that the BERT router is not significantly more efficient than a random strategy at identifying the most difficult GSM8K problems. This suggests that the general-purpose BERT router from RouteLLM \cite{Ong2024RouteLLM}, while effective across diverse domains, lacks the fine-grained discriminative power needed for highly specialized domains like mathematics. This highlights a potential area for future work in domain-specific router fine-tuning.

To address this limitation, several research directions can be explored. One direction is to enhance the router with fine-grained, domain-specific features, such as the number of variables or operations within a math problem. Another approach is domain-adaptive fine-tuning on a specialized dataset, though care must be taken to balance domain expertise with the potential degradation of the router's general-purpose capabilities. A third, more advanced direction involves using a powerful LLM as the router itself. For instance, Prompt-to-Leaderboard (P2L) trains a 7B model to generate prompt-specific leaderboards for optimal routing \cite{Frick2025P2L}. While potentially more accurate, such methods introduce significant routing overhead in terms of both latency and computation, which must be carefully balanced against the gains in overall system efficiency.

\begin{table}[t]
\centering
\caption{Baseline Model Performance on Multi-Turn Benchmark.}
\label{tab:multi_turn_perf}
\renewcommand{\arraystretch}{1.2}
\begin{tabular}{lcc}
\hline
\textbf{Model} & \textbf{Score} & \textbf{Resp. Len. (token)} \\
\hline
SLM (Qwen2.5-3B) & 9.29 & 226.37 \\
LLM (QwQ-32B) & 9.78 & 879.16 \\
\hline
\end{tabular}
\end{table}

\subsection{Multi-Turn Dialogue Routing Performance}
\begin{figure}[t]
\centering
\includegraphics[width=\columnwidth]{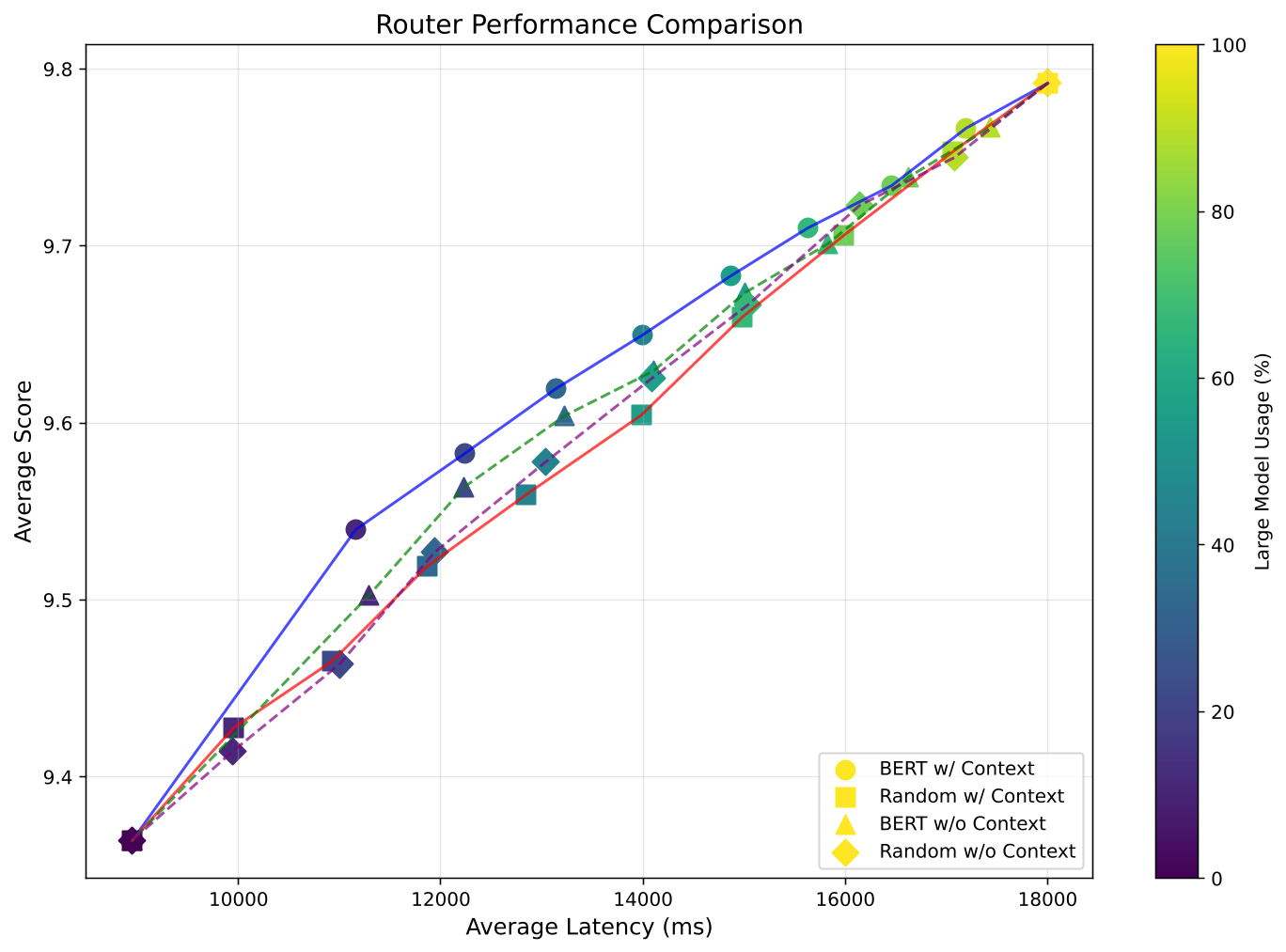}
\caption{Performance comparison on the MT-Bench-101 benchmark. The context-aware BERT router (``BERT w/ Context") provides the best tradeoff between average score and latency.}
\label{fig:mtbench_tradeoff}
\end{figure}

For the more complex multi-turn dialogue scenario, we assessed the combined impact of our routing logic and the context-aware cost model. The results on MT-Bench-101 are presented in Fig. \ref{fig:mtbench_tradeoff}. The graph reveals two key insights: (i) semantic-aware routing (BERT, solid lines) consistently outperforms the random baseline (dashed lines); (ii) context-aware modeling (``w/ Context") yields better performance than its context-agnostic counterpart (``w/o Context").

The benefits are most pronounced at a target score of approximately 9.53. At this performance level, our full \textit{BERT w/ Context} framework is the most efficient, requiring an average latency of only \textbf{11,157ms} with a lean \textbf{11.1\%} LLM usage rate. In contrast, removing context awareness (\textit{BERT w/o Context}) increases latency to 12,233ms and doubles the LLM usage to 22.2\%. The non-semantic baseline (\textit{Random w/ Context}) is even less efficient, needing 12,840ms of latency and a 44.4\% LLM usage rate to achieve a similar score. This detailed comparison demonstrates that our framework's advantages stem from both its semantic intelligence and its explicit modeling of context-related costs. This highlights the critical importance of jointly considering query semantics and KV-cache recomputation overhead in conversational settings.

\section{Conclusion}
In this paper, we propose and validate a dynamic routing framework to address the critical quality-latency trade-off in edge-device collaborative LLM inference. At its core, our framework's innovation is a cost model that, for the first time, makes dynamic routing practical in this challenging environment by co-optimizing for semantic difficulty and the critical KV-cache recomputation overhead in multi-turn dialogues. Experimental results on standard benchmarks confirm that our approach reduces average latency by 5-15\% and costly edge model invocations by 10-20\%, without compromising inference quality. This work presents a practical and effective pathway toward realizing responsive and resource-aware AI in future edge-device collaboration.

\bibliographystyle{IEEEtran}
\bibliography{references}{}
\vspace{12pt}

\end{document}